\documentclass[12pt, a4paper]{article} 
\usepackage{cosyne}
\usepackage{lipsum}
\usepackage{graphicx}
\usepackage{wrapfig}
\usepackage{caption}
\usepackage{placeins}
\usepackage{subcaption}
\title{\color{custom}\large{\vspace{-1.5cm}%
    Integrated representational signatures strengthen specificity in brains and models}}
\date{}
\author{
  Jialin Wu,
  Shreya Saha,
  Yiqing Bo,
  Meenakshi Khosla \\
  \texttt{\{jlwu,ssaha,ybo,mkhosla\}@ucsd.edu}
}

\begin{document}
\pagenumbering{gobble}
\maketitle
\vspace{-5.8em}
\textit{The extent to which different neural or artificial neural networks (models) rely on equivalent representations to support similar tasks remains a central question in neuroscience and machine learning. Prior work has typically compared systems using a single representational similarity metric, yet each captures only one facet of representational structure. To address this, we leverage a suite of representational similarity metrics—each capturing a distinct facet of representational correspondence, such as geometry, unit-level tuning, or linear decodability—and assess brain region or model separability using multiple complementary measures. Metrics that preserve geometric or tuning structure (e.g., RSA, Soft Matching) yield stronger region-based discrimination, whereas more flexible mappings such as Linear Predictivity show weaker separation. These findings suggest that geometry and tuning encode brain-region- or model-family-specific signatures, while linearly decodable information tends to be more globally shared across regions or models. To integrate these complementary representational facets, we adapt Similarity Network Fusion (SNF), a framework originally developed for multi-omics data integration. SNF produces substantially sharper regional and model family-level separation than any single metric and yields robust composite similarity profiles. Moreover, clustering cortical regions using SNF-derived similarity scores reveals a clearer hierarchical organization that aligns closely with established anatomical and functional hierarchies of the visual cortex—surpassing the correspondence achieved by individual metrics. }

\vspace*{-0.5em}
\noindent\rule{\columnwidth}{1pt}

\begin{figure}[ht]
  \centering
    \vspace{-8pt}
  \begin{minipage}[t]{0.79\linewidth}
    \setlength{\tabcolsep}{0pt}\renewcommand{\arraystretch}{1}
    \begin{tabular}{@{}m{1.6em}@{\hspace{0.4em}}m{\dimexpr\linewidth-1.6em-0.6em\relax}@{}}
      \textbf{A.} & \includegraphics[width=\linewidth]{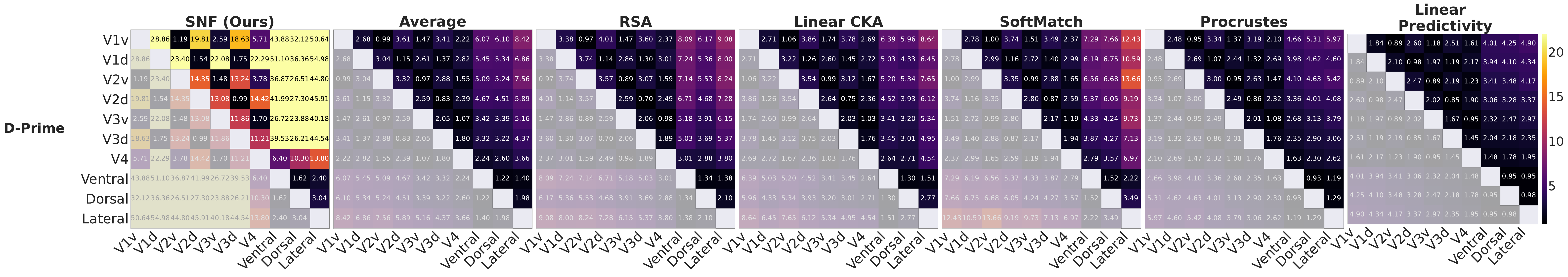}
    \end{tabular}
  \end{minipage}\hfill
  \begin{minipage}[t]{0.2\linewidth}
    \setlength{\tabcolsep}{0pt}\renewcommand{\arraystretch}{1}
    \begin{tabular}{@{}m{1.2em}@{\hspace{0.1em}}m{\dimexpr\linewidth-1.6em-0.6em\relax}@{}}
      \textbf{B.} & \includegraphics[width=\linewidth]{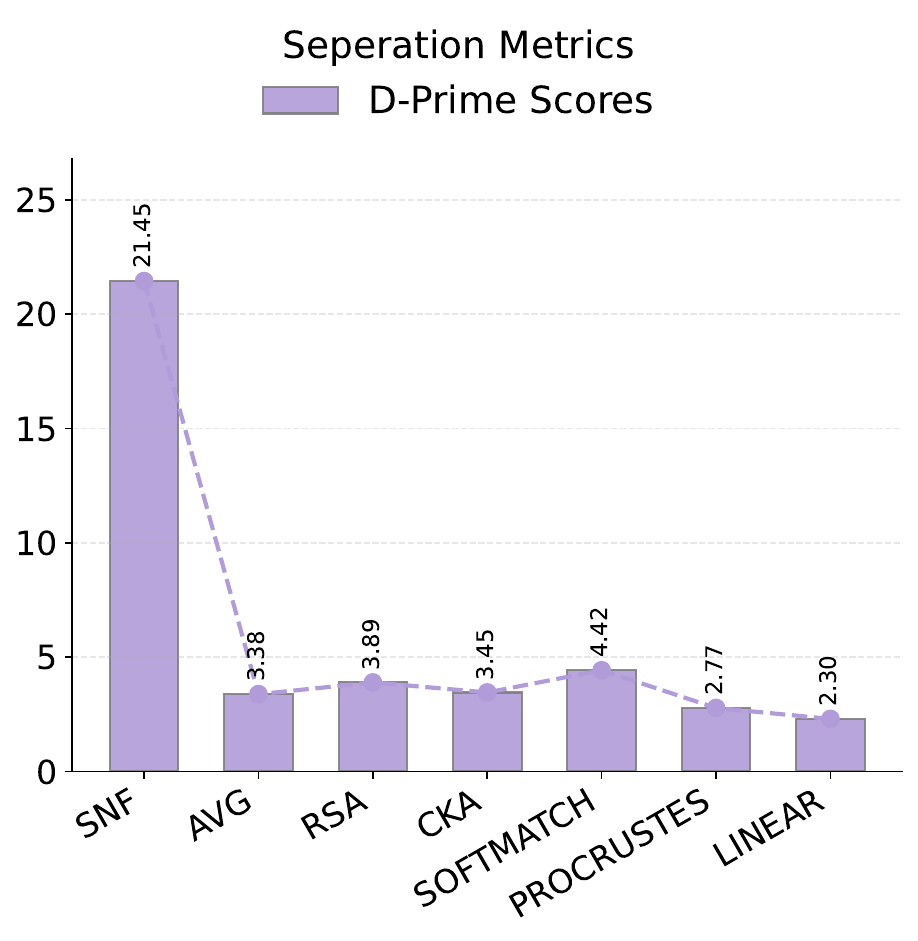}
    \end{tabular}
  \end{minipage}


  \begin{minipage}[t]{\linewidth}
    \setlength{\tabcolsep}{0pt}\renewcommand{\arraystretch}{1}
    \begin{tabular}{@{}m{1.2em}@{\hspace{0.1em}}m{\dimexpr\linewidth-1.6em-0.6em\relax}@{}}
      \textbf{C.} & \includegraphics[width=\linewidth]{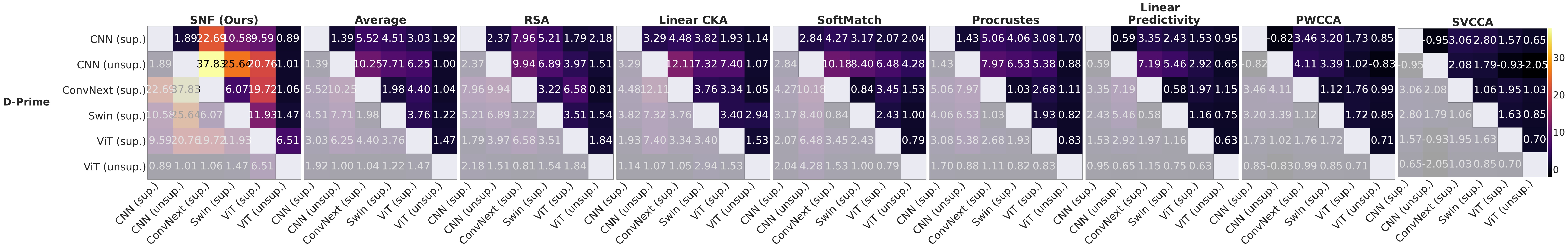}
    \end{tabular}
  \end{minipage}

  \vspace{-6pt}

  \caption{%
    \textbf{A.} Brain region separability under $d'$. Columns correspond to seven similarity metrics, including two fusion-based methods (SNF, average) and five commonly used representational metrics. The color bar is capped at 22. 
    \textbf{B.} Mean separability score on NSD. Scores are shown in their native scales.
    \textbf{C.} Same analysis as in A, applied to vision model families. }
  \label{fig:abc}
  \vspace{-12pt}
\end{figure}

Comparisons of neural representations—across individuals, brain regions, or model instances—typically rely on a single representational similarity metric. However, each metric typically captures only one facet of representational structure, such as representational geometry, unit tuning, or linearly accessible information, leaving other dimensions unaccounted for~\citep{rsa,yamins2014performance, kornblith2019similarity, khosla2024soft}. Integrating these complementary signatures could yield comparisons that more faithfully reflect underlying representational similarity and offer more interpretable insights into shared mechanisms. A key desideratum for any representational comparison is specificity—the extent to which it correctly identifies shared mechanisms while distinguishing distinct ones~\citep{thobani2025modelbraincomparisonusinginteranimal}. Specifically, a good metric should strongly align responses from the same brain area across subjects (or the same model family) while maintaining clear separation between different areas or families. We use this criterion to evaluate existing metrics and introduce an integrated representational comparison framework, inspired by multi-omics research, based on Similarity Network Fusion (SNF)~\citep{wang2014similarity,snfpy}. This approach achieves substantially higher specificity across both neural and model representations. 

In SNF, we first convert the similarity matrix $S^v$ obtained from each metric $v$ into symmetricized distances: $Q^v_{ij}=\mathbf{1}_{i\neq j}\bigl(1-({S^v_{ij}+S^v_{ji}})/{2}\bigr)$.
We then build an affinity $\mathbf{W}^v$ using a Gaussian kernel with bandwidth $\sigma^v_{ij}=\mu\,({\bar{Q}^v(i,N_i)+\bar{Q}^v(j,N_j)+Q^v_{ij}})/{3}$, where $\bar{Q}^v(i,N_i)$ is the mean dissimilarity from $i$ to its $K = 5$ nearest neighbors $N_i$, and $\mu = 0.5$.
Then we do row normalization and symmetricization and get $\widehat{\mathbf{W}}^{\,v}$. Subsequently, we perform iterative message passing on each graph. We initialize $\mathbf{P}^{(v)}_{0}=\widehat{\mathbf{W}}^{\,v}$; for $t=0,\dots,T-1$ let $\mathbf{P}^{(v)}_{t+1}=\mathcal{B}_\alpha\!\Big(\mathbf{S}^{(v)}\big(\tfrac{1}{|\mathcal{V}|-1}\sum_{u\neq v}\mathbf{P}^{(u)}_{t}\big)\mathbf{S}^{(v)\!\top}\Big)$, with $\mathcal{B}_\alpha(\mathbf{X})=(\mathbf{X}+\mathbf{X}^\top)/2+\mathbf{I}$ and $\mathbf{S}^{(v)}_{ij}=\widehat{\mathbf{W}}^{\,v}_{ij}/\sum_{k\in N_i}\widehat{\mathbf{W}}^{\,v}_{ik}$ if $j\in N_i$, else $0$.
After $T=20$ rounds, the graphs are averaged, followed by row normalization and symmetrization, yielding a fused similarity matrix that highlights cross-metric consistent relationships while suppressing noise.

For brain data, we used responses to the 1,000 shared natural images from the Natural Scenes Dataset (NSD)~\citep{nsd}.
For models, our stimulus set is the ImageNet validation set~\citep{imagenet_cvpr09}. We analyzed 35 ImageNet-trained models across 6 primary families: supervised CNNs, self-supervised CNNs, supervised Transformers, self-supervised Transformers, and the hybrid architectures ConvNeXt~\cite{convnet} and Swin~\cite{swin}, treated as distinct categories due to their CNN-Transformer design.

We compared representational similarity using RSA~\citep{rsa}, Linear CKA~\citep{kornblith2019similarity}, Soft Matching (SoftMatch)~\citep{khosla2024soft}, Procrustes~\citep{ding2021grounding}, and Linear Predictivity~\citep{yamins2014performance}.
For models, we additionally included SVCCA~\citep{raghu2017svcca} and PWCCA~\citep{morcos2018insights}, which were excluded from brain analyses because their alignment values were inflated by the high voxel count relative to the number of stimuli. As a baseline, we symmetrized, min-max rescaled, and averaged all metric matrices to form a naïve aggregate similarity estimate.

To assess brain region discriminability, we quantified the difference between within-region and across-region representational similarities using the separability measure $d'$. SNF which integrates information across all representational dimensions, achieves dramatically superior family separation compared to any single metric. In Figure~\ref{fig:abc}B, SNF attains a mean $d'$ of 21.45—nearly five times higher than the best-performing single measure—and consistently outperforms all baselines across separation criteria. Importantly, in Figure~\ref{fig:abc}A, SNF maintains high and balanced discrimination across nearly all region pairs. By contrast, individual metrics often exhibit uneven performance, separating some regions while failing for others.
Averaging similarities across metrics does not resolve this limitation: simple means dilute complementary signals and retain conflicting noise. In contrast, SNF’s diffusion-based fusion reinforces consistent neighborhood structure across metrics while attenuating discordant components, yielding both stronger global separation and greater local stability.

\begin{wrapfigure}{r}{0.36\textwidth}
    \vspace{-12pt}
    \centering
    \includegraphics[width=\linewidth]{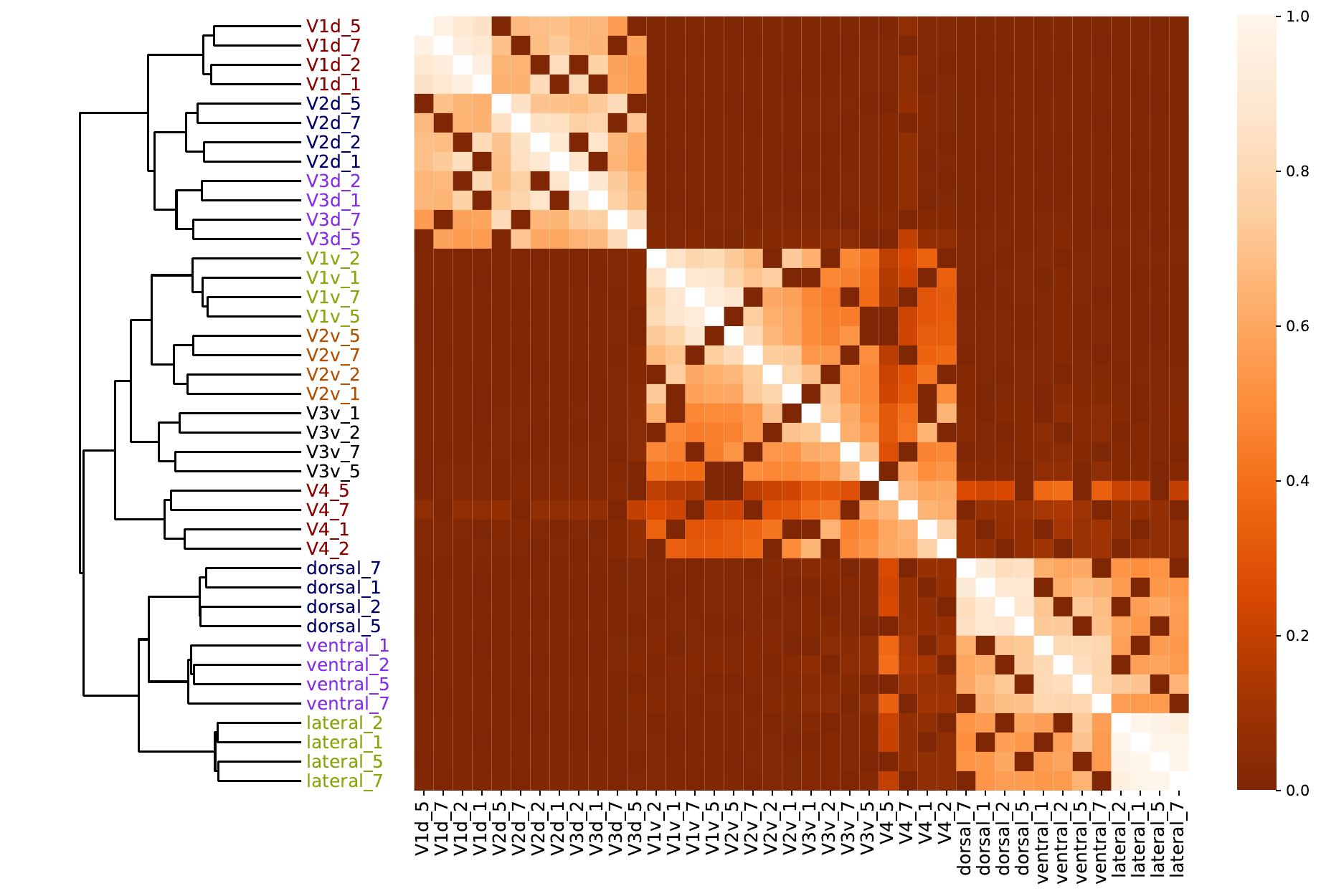}
    \caption{The heatmap shows the SNF-fused similarity matrix reordered by leaf ordering. Leaf labels are formatted as $region\_subject$ and colored by the cluster they belong to; dendrogram cuts yield up to ten flat clusters aligned with canonical categories. Considering high correlations across regions of the same subject caused by fMRI property, we zeroed out these similarity values to exclude the subject bias.}
    \vspace{-5pt}
    \label{fig:snf_clustering_NSD}
    \vspace{-22pt}
\end{wrapfigure}

Beyond scalar separability, clustering the SNF-fused similarity matrix exposed clear cross-subject structure: responses from the same cortical area formed tight, subject-invariant clusters (Figure~\ref{fig:snf_clustering_NSD}).
At a broader scale, the SNF-fused matrix uncovered two major superclusters: one spanning the early visual cortex (V1-V4) and another comprising the high-level visual streams (ventral, dorsal, and lateral). Within the early cluster, the dorsal and ventral subdivisions of V1, V2, and V3 clustered separately, consistent with known anatomical organization.

\begin{wrapfigure}{r}{0.54\textwidth}
    \vspace{-14pt}
    \centering
    \includegraphics[width=\linewidth]{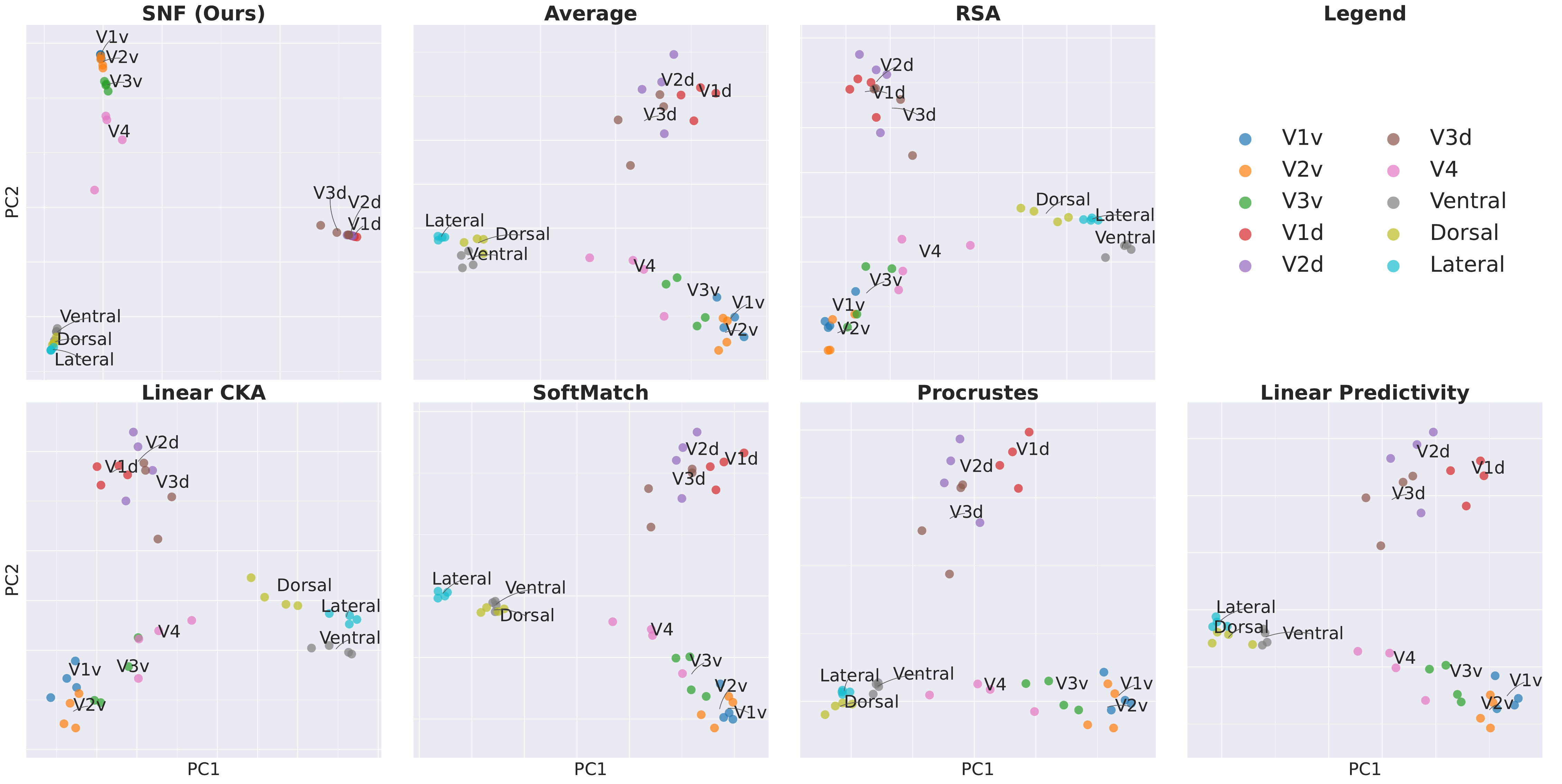}
    \caption{Cross-regional relationships derived from five similarity measures using PCA analysis on NSD data. Each point represents a brain region instance, and text labels indicate centroid positions. SNF fusion shows best intra-class compactness and inter-class separation.}
    \label{fig:pca}
    \vspace{-16pt}
\end{wrapfigure}

To visualize and corroborate these patterns, we applied PCA to each similarity matrix. All metrics recovered broad trends (Fig. \ref{fig:pca}), but PCA of the SNF-fused matrix produced the clearest, most interpretable layout. Early visual areas (V1-V4) traced a smooth trajectory, with V4 in a transitional position between early cortex and the higher-level ventral, dorsal, and lateral streams. In contrast, RSA, Linear CKA, and SoftMatch yielded compressed or overlapping clusters, while the other two produced more diffuse arrangements. Overall, SNF best preserved both within-area compactness and between-area separation, consistent with known anatomy.

For artificial neural networks, SNF likewise achieves markedly superior performance in capturing within-family identifiability while maintaining between-family separability (Figure~\ref{fig:abc}C). It yields high and balanced discrimination across nearly all family pairs, whereas individual metrics show uneven performance and the averaged baseline fails to resolve this limitation. Metrics emphasizing geometry and tuning capture family-specific computational signatures, whereas those emphasizing linear mappings blur distinctions across architectures or training paradigms. By integrating these complementary facets, SNF produces composite representational signatures that most reliably distinguish model families.

These results highlight how integrating complementary representational facets can uncover the organizing principles that unify—and distinguish—neural or artificial systems.

\printbibliography

\end{document}